\documentclass[finnish,english,aip,jcp,reprint]{revtex4-1}
\usepackage[T1]{fontenc}
\usepackage[utf8]{inputenc}
\usepackage{refstyle}
\usepackage{amsmath}
\usepackage{graphicx}

\makeatletter

\AtBeginDocument{\providecommand\figref[1]{\ref{fig:#1}}}
\AtBeginDocument{\providecommand\eqref[1]{\ref{eq:#1}}}
\AtBeginDocument{\providecommand\tabref[1]{\ref{tab:#1}}}
\providecommand{\tabularnewline}{\\}
\RS@ifundefined{subsecref}
  {\newref{subsec}{name = \RSsectxt}}
  {}
\RS@ifundefined{thmref}
  {\def\RSthmtxt{theorem~}\newref{thm}{name = \RSthmtxt}}
  {}
\RS@ifundefined{lemref}
  {\def\RSlemtxt{lemma~}\newref{lem}{name = \RSlemtxt}}
  {}

\usepackage{chemfig}
\usepackage[version=3]{mhchem}
\usepackage{braket}

\@ifundefined{showcaptionsetup}{}{%
 \PassOptionsToPackage{caption=false}{subfig}}
\usepackage{subfig}
\makeatother

\usepackage{babel}
\begin{document}

\title{Cost-effective description of strong correlation: efficient implementations
of the perfect quadruples and perfect hextuples models}

\author{Susi Lehtola}

\affiliation{Chemical Sciences Division, Lawrence Berkeley National Laboratory,
Berkeley, California 94720, United States}
\email{susi.lehtola@alumni.helsinki.fi}

\author{John Parkhill}

\affiliation{Department of Chemistry, University of Notre Dame, 251 Nieuwland
Science Hall, Notre Dame, Indiana 46556, United States}
\email{john.parkhill@gmail.com}

\author{Martin Head-Gordon}

\affiliation{Chemical Sciences Division, Lawrence Berkeley National Laboratory,
Berkeley, California 94720, United States}

\affiliation{Department of Chemistry, University of California, Berkeley, California
94720, United States}
\email{mhg@cchem.berkeley.edu}

\begin{abstract}
Novel implementations based on dense tensor storage are presented
for the singlet-reference perfect quadruples (PQ) {[}Parkhill, Lawler,
and Head-Gordon, J. Chem. Phys. 130, 084101 (2009){]} and perfect
hextuples (PH) {[}Parkhill and Head-Gordon, J. Chem. Phys. 133, 024103
(2010){]} models. The methods are obtained as block decompositions
of conventional coupled-cluster theory that are exact for four electrons
in four orbitals (PQ) and six electrons in six orbitals (PH), but
that can also be applied to much larger systems. PQ and PH have storage
requirements that scale as the square, and as the cube of the number
of active electrons, respectively, and exhibit quartic scaling of
the computational effort for large systems. Applications of the new
implementations are presented for full-valence calculations on linear
polyenes (\ce{C_nH_{n+2}}), which highlight the excellent computational
scaling of the present implementations that can routinely handle active
spaces of hundreds of electrons. The accuracy of the models is studied
in the $\pi$ space of the polyenes, in hydrogen chains (\ce{H50}),
and in the $\pi$ space of polyacene molecules. In all cases, the
results compare favorably to density matrix renormalization group
values. With the novel implementation of PQ, active spaces of 140
electrons in 140 orbitals can be solved in a matter of minutes on
a single core workstation, and the relatively low polynomial scaling
means that very large systems are also accessible using parallel computing.
\end{abstract}
\maketitle
\selectlanguage{finnish}%
\global\long\def\ERI#1#2{(#1|#2)}
\foreignlanguage{english}{}\global\long\def\bra#1{\Bra{#1}}
\foreignlanguage{english}{}\global\long\def\ket#1{\Ket{#1}}
\foreignlanguage{english}{}\global\long\def\braket#1{\Braket{#1}}

\selectlanguage{english}%

\section{Introduction}

Efficient handling of strong correlation is one of the still unresolved
questions in computational chemistry. Outside the scope of the otherwise
useful Kohn–Sham density-functional theory\citep{Hohenberg1964,Kohn1965,Kurzweil2009},
approaches to strong correlation lean heavily on wave function theories.
The conventional way to treat strong correlation is through multiconfigurational
self-consistent field (MCSCF) theory, which scales exponentially as
the size of the active space grows. As a result, MCSCF cannot be applied
to active spaces significantly larger than 16 electrons in 16 orbitals,
although the barrier has been recently pushed back through approximative
stochastic\citep{Thomas2015,LiManni2016} and adaptive\citep{Schriber2016,Tubman2016,Liu2016,Zhang2016a,Holmes2016}
approaches to the full configuration interaction (FCI) problem.

Now, while MCSCF relies on a configuration interaction (CI) type ansatz
$\ket{\Psi}=(1+\hat{T})\ket{\Phi}$ for the wave function, where $\ket{\Phi}$
is the single-particle reference and $\ket{\Psi}$ the true, multiconfigurational
wave function, one might ask whether a coupled-cluster\citep{Cizek1966}
(CC) type ansatz $\ket{\Psi}=\exp(\hat{T})\ket{\Phi}$ could be used
instead to formulate a MCSCF-type model. The CC approach can also
be used to describe dynamic correlation with a CI reference, as in
the multireference CC (MR-CC) approach\citep{Lyakh2012}, but MR-CC
is much more complicated than single-reference CC, and combining both
strong correlation and dynamic correlation within the same machinery
would be clearly advantageous. While in principle CC theory is well
able to describe static correlation, the problem that arises upon
its application is that the computational effort of the resulting
procedure is prohibitively expensive, even if an active space is used\citep{Krylov1998}.
However, this problem has been recently solved by Parkhill, Lawler,
and Head-Gordon, who proposed a truncation of CC for exactness within
the active space for a given number of strongly interacting electron
pairs\citep{Parkhill2009,Parkhill2010}, instead of the usual, global
truncation based on the excitation level. For example, the perfect
quadruples\citep{Parkhill2009} (PQ) model is formed by truncating
CC theory with single through quadruple (CCSDTQ) excitations to two
pairs by keeping only the amplitudes $t_{i_{1}\dots i_{n}}^{a_{1}\dots a_{n}}$
that satisfy $\{a_{k},i_{k}\}\in\{\text{pair}_{1}\}\times\{\text{pair}_{2}\}$.
Similarly, the perfect hextuples\citep{Parkhill2010} (PH) model is
a three-pair truncation of CC theory with single through hextuple
excitations (CCSDTQ56), in which only the amplitudes that satisfy
$\{a_{k},i_{k}\}\in\{\text{pair}_{1}\}\times\{\text{pair}_{2}\}\times\{\text{pair}_{3}\}$
are kept. Here, the concept of a \emph{pair} arises from connection
to the perfect pairing\citep{Hurley1953,Hunt1972,Ukrainskii1977,Goddard1978,Cullen1996}
(PP) theory, in which pair $i$ is described by a quartet of orbitals:
the alpha and beta orbitals $i$ and $\bar{i}$, respectively, that
are occupied in the single-particle reference, as well as the corresponding
alpha and beta virtual orbitals $i^{*}$ and $\bar{i}^{*}$, respectively.
In contrast to MCSCF, PQ and PH exhibit polynomial scaling with respect
to the size of the active space. The storage requirements are modest,
scaling as $O(N^{2})$ for PQ and $O(N^{3})$ for PH, $N$ being the
number of active orbitals in the calculation. The computational effort
scales asymptotically as $O(N^{4}$) for both models.

Another widely used approach for strong correlation is the density
matrix renormalization group (DMRG) method\citep{White1992,Chan2008,Marti2010,Wouters2013,Olivares-Amaya2015},
which is exact in principle and which also offers polynomial scaling
with respect to system size. However, the method has a significant
prefactor: the storage and computational costs scale as\citep{Hachmann2006,Olivares-Amaya2015}
$O(M^{2}N^{3})$ and $O(M^{3}N^{3})+O(M^{2}N^{4})$, respectively,
where $M$ is the number of states kept in the calculation. Furthermore,
as a fundamentally one-dimensional approach, DMRG has a history of
being hard to use due to issues with orbital ordering, although automatic
orbital ordering schemes have recently made DMRG calculations easier
to perform. We note that other approaches for a compact description
of strong correlation have also been suggested, such as the CC valence
bond (CCVB) method\citep{Small2009,Small2011,Small2014}, extended
CC\citep{Arponen1983,Arponen1987,Arponen1987a}, and the multifacet
graphically contracted function method\citep{Shepard2014,Shepard2014a}.

In order to apply MCSCF, DMRG, PQ, or PH to studies of chemical systems,
the additional description of dynamic correlation is usually of utmost
importance. While MCSCF and DMRG typically rely on a two-pronged approach
to dynamic correlation through CI, CC, or perturbative approaches,
dynamic correlation can be treated on the same footing as static correlation
in PQ and PH\citep{Parkhill2010b}, enabling the two types of correlation
to interact. Note that perturbative approaches can be used as well
in combination with PQ\citep{Parkhill2011} and PH.

What unites MCSCF, DMRG, PQ, and PH is that the choice of the set
of active orbitals is a problem, and often the orbitals need to be
optimized in order to fully capture static correlation effects. Orbital
optimization is hard for PQ and PH, but earlier experience\citep{Parkhill2010}
suggests that in many cases orbitals optimized at the much simpler
PP level of theory might be sufficient for a wide variety of applications.
Thus, in the present work, only localized Hartree–Fock orbitals and
PP optimized orbitals will be considered. Orbitals optimized with
CCVB, or unrestricted Hartree–Fock natural orbitals\citep{Pulay1988,Bofill1989,Keller2015}
might also prove useful. However, because the feasible size of the
active space is much larger in PQ and PH than in MCSCF or even DMRG,
it is likely that for full-valence calculations the choice of the
active space orbitals should not prove to be an impediment to the
use of the PQ and PH models.

While encouraging results have been achieved withPQ and PH, their
existing implementations (while exhibiting the correct $O(N^{4})$
scaling with system size) have a too large prefactor to allow their
application to studies of chemical problems. Because the previous
implementations are based on the use of sparse tensor algebra\citep{Parkhill2010a},
it is worth asking if the prefactor could be reduced by the use of
dense tensor storage, instead. Moreover, the existing sparse tensor
implementation does not take advantage of the mathematical structure
of the PQ and PH methods, which results in extra work for bookkeeping
and prevents efficient compiler optimization from taking place.

In the present work, we describe the dense tensor implementation of
the PQ and PH models, and benchmark the resulting code against the
previous implementation on linear polyenes. Novel applications of
the PQ and PH models are demonstrated for the \ce{H50} chain, and
the results are compared to DMRG. Applications of PQ and PH to the
strongly correlated $\pi$ space of the polyacene series are also
presented. The organization of the manuscript is the following. In
the Theory section, we illustrate how the truncation of CC theory
is performed. Next, in the Implementation section, we describe various
aspects of how the procedure has been carried out in practice. In
the Results section we present the scaling of the new code when applied
to linear polyenes, and the energies for the polyenes, the dissociation
of the \ce{H50} chain, as well as the polyacenes for which natural
occupation numbers are also presented. The study concludes with a
summary and brief discussion.

\section{Theory}

\begin{figure}
\begin{centering}
\includegraphics[width=6cm]{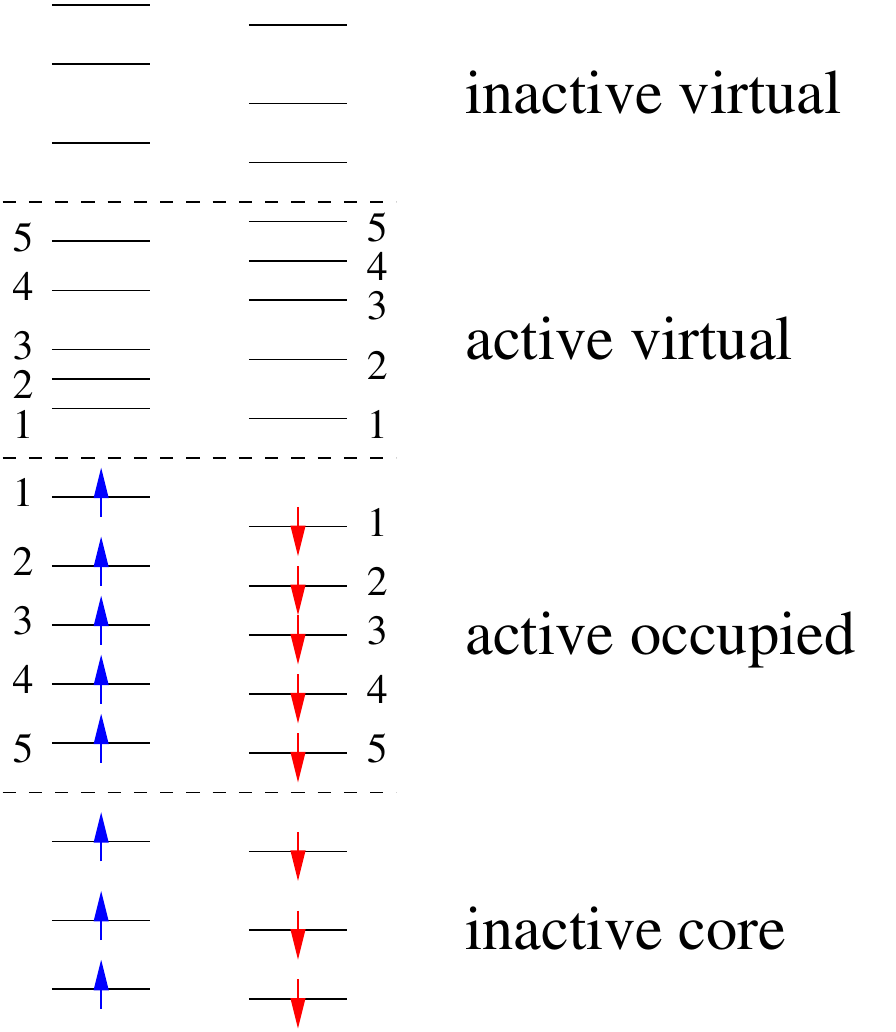}
\par\end{centering}
\caption{Division of the orbital space. The active space orbitals are numbered
with the value of the pairing label. The hole indices are reversed,
because the strongest coupling typically occurs between the highest
occupied and lowest unoccupied orbitals.\label{fig:Division-of-the}}
\end{figure}

The theory behind the CC approach is well established and shall not
be discussed further here\citep{Crawford2000,Bartlett2007}. The starting
point for the present truncation are the CC equations in spin-integrated
form, consisting of tensors that belong to a certain spin block, such
as the $\alpha\beta$ block of the double excitation amplitudes, where
$\alpha$ and $\beta$ denote spin-up and spin-down, respectively.
As a truncation of CC theory, the models in the present hierarchy
inherit all the beneficial properties of CC such as size extensivity,
but also its drawbacks such as non-variational behavior in cases where
the excitation rank in CC is not high enough to describe the static
correlation effects.

In the following we will omit the spin indices, because the discussion
applies to all tensors that appear in CC theory. We will mostly be
focusing on the PQ model, as it already illustrates all of the necessary
aspects for a general truncation method without being overly complicated.
The division of the orbital space that is used in the present work
is illustrated in \figref{Division-of-the}. The system, in a singlet
reference state, is composed of a set of inactive core orbitals, a
set of $N$ active occupied orbitals, a set of $N$ active virtual
orbitals, and a set of inactive virtual orbitals. The tensors, such
as the excitation amplitudes $t$, the two-electron integrals $v$,
the de-excitation amplitudes $\lambda$ and the one- and two-electron
density matrices $\gamma$ and $\Gamma$ are truncated using a block
decomposition. For instance, for PQ it is not hard to see that the
expansion of the double excitation tensor as
\begin{widetext}
\begin{align}
t_{ij}^{ab}= & \sum_{P=1}^{N}t_{PP}^{PP}\delta_{iP}\delta_{jP}\delta_{aP}\delta_{bP}+\sum_{\substack{P,Q=1\\
P\neq Q
}
}^{N}t_{PP}^{PQ}\delta_{iP}\delta_{jP}\delta_{aP}\delta_{bQ}+\sum_{\substack{P,Q=1\\
P\neq Q
}
}^{N}t_{PP}^{QP}\delta_{iP}\delta_{jP}\delta_{aQ}\delta_{bP}+\sum_{\substack{P,Q=1\\
P\neq Q
}
}^{N}t_{PQ}^{PP}\delta_{iP}\delta_{jQ}\delta_{aP}\delta_{bP}\label{eq:decompose}\\
+ & \sum_{\substack{P,Q=1\\
P\neq Q
}
}^{N}t_{QP}^{PP}\delta_{iQ}\delta_{jP}\delta_{aP}\delta_{bP}+\sum_{\substack{P,Q=1\\
P\neq Q
}
}^{N}t_{PQ}^{PQ}\delta_{iP}\delta_{jQ}\delta_{aP}\delta_{bQ}+\sum_{\substack{P,Q=1\\
P\neq Q
}
}^{N}t_{QP}^{PQ}\delta_{iQ}\delta_{jP}\delta_{aP}\delta_{bQ}+\sum_{\substack{P,Q=1\\
P\neq Q
}
}^{N}t_{PP}^{QQ}\delta_{iP}\delta_{jP}\delta_{aQ}\delta_{bQ}\nonumber 
\end{align}
\end{widetext}
where $i$ and $j$ denote occupied orbitals, $a$ and $b$ denote
virtual orbitals, $P$ and $Q$ are pairing labels that run over the
active space and $\delta_{PQ}$ is the Kronecker delta is equivalent
to the original definition of the PQ model as the set of amplitudes
$t_{i_{1}\dots i_{n}}^{a_{1}\dots a_{n}}$ that satisfy $\{a_{k},i_{k}\}\in\{\text{pair}_{1}\}\times\{\text{pair}_{2}\}$.
As this kind of notation is unusual to most chemists, we will try
to clarify the meaning of \eqref{decompose} in the following. In
the convention adapted above, all the orbitals in the alpha and beta
occupied and virtual spaces are numbered from 1 to $N$, which is
also the range spanned by the pairing labels. Because we have assumed
that the spin has been integrated out, all the quantities in \eqref{decompose}
are scalar numbers. The indices $i,j,a,b$ are fixed by the left hand
side of the equation, which represents the element of the excitation
tensor under decomposition. The Kronecker delta is defined as $\delta_{ij}=1$
when $i=j$ and $\delta_{ij}=0$ when $i\neq j$. Applying the decomposition
to, for example, the element $t_{kk}^{cc}$ with $k\neq c$ will only
yield a contribution from the last term of \eqref{decompose} (which
equals $t_{kk}^{cc}$), because the first term would require $k=c$
and all the other terms would contribute only for $P=Q$ which has
been excluded in the summations.

The restriction to identically sized active occupied and virtual spaces
becomes apparent here, as the same pairing labels cannot be used for
spaces of different size. Each value of a pairing label corresponds
to a quartet of orbitals: the alpha and beta occupied orbital, and
the alpha and beta virtual orbital. The tensor decomposition in \eqref{decompose}
is known in mathematics as a block decomposition\citep{DeLathauwer2008a,DeLathauwer2008}.
For PH, in addition to the 8 cases shown in \eqref{decompose} one
obtains 63 more, corresponding to all the inequivalent ways of distributing
three labels to four indices: for instance, $t_{QQ}^{QP}$ is equivalent
to $t_{PP}^{PQ}$ as the dummy summation indices $P$ and $Q$ can
be interchanged. Having enumerated the different subtensor representations
for all the necessary tensors, one proceeds by inserting them into
the CC equations and by performing the summations over the Kronecker
deltas. When this is done, one ends up with a formulation of PQ or
PH theory in terms of the dense subtensors. For example, in the case
of PH the original hextuple excitation operator $T_{ijklmn}^{abcdef}$,
which is a rank-12 tensor, is reduced to a set of rank-3 tensors,
and any summations that are left over are only with respect to the
pairing labels. While in principle, the doubles contribution to the
opposite spin correlation energy 
\begin{align}
E_{c}\leftarrow & \sum_{ijab}t_{ij}^{ab}v_{ij}^{ab}\label{eq:Ec}
\end{align}
has $8\times8=64$ distinct subtensor contributions in PQ (and 5041
in PH), it is seen that due to orthogonality of the Kronecker indices
the PQ expression becomes simply
\begin{widetext}
\begin{align}
E_{c}\leftarrow & \sum_{P}t_{PP}^{PP}v_{PP}^{PP}+\sum_{\substack{P,Q=1\\
P\neq Q
}
}^{N}t_{PP}^{PQ}v_{PP}^{PQ}+\sum_{\substack{P,Q=1\\
P\neq Q
}
}^{N}t_{PP}^{QP}v_{PP}^{QP}+\sum_{\substack{P,Q=1\\
P\neq Q
}
}^{N}t_{PQ}^{PP}v_{PQ}^{PP}\label{eq:energy-contr}\\
+ & \sum_{\substack{P,Q=1\\
P\neq Q
}
}^{N}t_{QP}^{PP}v_{QP}^{PP}+\sum_{\substack{P,Q=1\\
P\neq Q
}
}^{N}t_{PQ}^{PQ}v_{PQ}^{PQ}+\sum_{\substack{P,Q=1\\
P\neq Q
}
}^{N}t_{QP}^{PQ}v_{QP}^{PQ}+\sum_{\substack{P,Q=1\\
P\neq Q
}
}^{N}t_{PP}^{QQ}v_{PP}^{QQ}.\nonumber 
\end{align}
\end{widetext}
The limitation to $P\neq Q$ in \eqref{energy-contr} and similar
equations can be easily lifted by establishing a convention in which
all diagonals of the subtensors vanish.

One can identify a set of 8 subtensors in \eqref{decompose,energy-contr}:
$t_{PP}^{PP}(P)$, $t_{PP}^{PQ}(P,Q)$, $t_{PP}^{QP}(P,Q)$, $t_{PQ}^{PP}(P,Q)$,
$t_{QP}^{PP}(P,Q)$, $t_{PQ}^{PQ}(P,Q)$, $t_{QP}^{PQ}(P,Q)$, and
$t_{PP}^{QQ}(P,Q)$. The first term in \eqref{decompose}, $t_{PP}^{PP}(P)$
can be identified as the PP amplitude (singles are usually not included
in the definition of PP as the orbitals are normally optimized). As
the simplest truncation (exactness for two electrons in two orbitals),
PP has a special place within the hierarchy of the truncated models.
Because PP couples every occupied orbital to a specific virtual orbital,
the amplitudes are independent and can be solved for in closed form.
Orbital optimization can then be done in a routine fashion even for
large systems\citep{Beran2005}. The terms $t_{PP}^{PP}(P)$, $t_{PQ}^{PQ}(P,Q)$,
and $t_{QP}^{PQ}(P,Q)$ in \eqref{decompose} are the basis of the
imperfect pairing (IP) model\citep{VanVoorhis2000a}, which is considerably
more complicated than PP but has still been originally been derived
by hand, and for which orbital optimization is still a routine task. 

However, in addition to the terms in \eqref{decompose} that represent
double excitations, PQ also includes single, triple, and quadruple
excitations, and unlike PP and IP is not restricted to opposite-spin
correlation, so the derivation of the equations for PQ would be a
challenging task by hand. Fortunately, it is an ideal task for a computer.

Antisymmetricity between same-spin indices can be used to eliminate
redundant storage of equivalent contributions (\emph{e.g.,} $t_{PQ}^{QP}=-t_{PQ}^{PQ}$
for the $\alpha\alpha$ and $\beta\beta$ blocks in \eqref{decompose}).
The resulting storage costs for the subtensors that appear at the
PP, PQ, and PH levels of theory are shown in \tabref{Storage-costs-for}.
For instance, for PQ the storage cost is $17N^{2}+3N$ for the amplitudes,
$113N^{2}+19N$ for the density matrices and $115N^{2}+23N$ for the
integrals. The figures in \tabref{Storage-costs-for} also include
the classes of integrals that are only needed for the calculation
of the CC pseudoenergy; for a $t$ amplitude-only program the amount
of integrals would be slightly smaller. \Tabref{noamps} details the
distribution of the total amplitude subtensors into the individual
excitation ranks.

Due to the truncation of the amplitudes and integrals, PP, PQ, and
PH are not invariant to rotations in the occupied–occupied or virtual–virtual
blocks, unlike their parent CC theories. The same will also apply
perfect octuples (PO) theory, which would be a truncation of CC theory
with single through octuple excitations (CCSDTQ5678) based on exactness
for eight electrons in eight orbitals. This is somewhat counterintuitive,
because in contrast to PP, PQ, and PH, the full set of one- and two-electron
integrals is used in PO, encoding in principle all there is to know
about the system. But, the triple through octuple excitation operators
would still not be fully described within PO, and the truncation is
based on orbital labels, so the used set of orbitals would still be
an issue in PO. However, seeing as the issues with the choice of orbitals
clearly ease up going from PP to PQ to PH, PO might not be very sensitive
to the orbitals after all.

Due to the one-to-one coupling of occupied and virtual orbitals PP
(and IP) is obviously dependent on the relative ordering of the occupied
and virtual orbitals. But, the same property also applies to the PQ
and PH models. For instance, if one interchanges the virtual orbitals
2 and 3, the PQ doubles amplitude $t_{12}^{12}$ in the original numbering
would become $t_{12}^{13}$ in the renumbered system. However, $t_{12}^{13}$
is not included in the PQ model, so the description of the system
has changed. By a similar analysis, it is easy to see that the dependence
on the orbital orderings holds for all tensors that are truncated
in the subtensor expansion. Due to this common property, we refer
to the hierarchy of the PP, PQ, and PH (and higher models) as the
\emph{perfect pairing hierarchy} to underline the pairing of the orbital
quartets, formed of occupied and virtual alpha and beta orbitals,
at each level of theory.

The origin of the dependence on the relative ordering of the occupied
and virtual orbitals is easy to understand on physical grounds. In
order to reduce the computational scaling, we wished to retain only
the strongest interactions between orbitals. The pairing of occupied
orbitals to virtual orbitals introduces a sense of locality in the
model that results in fast convergence of the truncation, as in the
usual local dynamic correlation methods\citep{Pulay1983,Saebo1993}.
The models in the PP hierarchy are, however, invariant to swaps between
pairs (\emph{i.e.} quartets) of orbitals.

The models in the PP hierarchy assume a pairing of the occupied and
virtual orbitals before the calculation is began. While PP optimized
orbitals will produce the best possible pairing in the sense of reproducing
the lowest energy at the PP level of theory, it is not clear that
the PP orbitals, which usually turn out to be local, will be optimal
for the PQ or PH models: for instance in polyacenes, the optimal orbitals
for CCVB (which describes more strong correlation than PP) turn out
to be more delocalized than PP orbitals\citep{Small2014}. However,
as stated in the Introduction, variational optimization of the orbitals
for PQ and PH is hard, which means that one may need to resort to
approximate methods of obtaining localized orbitals with matching
virtual orbitals. But, this may not be a problem after all: while
orbital optimization is necessary at the PP level of theory due its
aggressive truncation (which also makes the optimization problem tractable),
it is plausible that the use of optimal orbitals is less important
at higher levels of the hierarchy, because at every step the models
become closer to FCI which is orbital invariant. In analogy to local
dynamic correlation methods, reasonable orbitals can be obtained from
a Hartree–Fock or density-functional theory calculation by localizing
the occupied orbitals using, \emph{e.g.}, the Foster–Boys\citep{Foster1960}
(FB), Edmiston–Ruedenberg\citep{Edmiston1963} (ER), or Pipek–Mezey\citep{Pipek1989,Lehtola2014}
(PM) criteria, after which matching virtual orbitals can be obtained,
\emph{e.g.}, by using the Sano procedure\citep{Sano2000}. In the
present work, both PP orbitals and Pipek–Mezey orbitals are used.

\begin{table}
\begin{centering}
\begin{tabular}{cccc}
 & rank-1 & rank-2 & rank-3\tabularnewline
\hline 
\hline 
$t$/$\lambda$ & 3 & 17 & 64\tabularnewline
$f$ & 10 & 6 & 0\tabularnewline
$v$ & 13 & 109 & 96\tabularnewline
$\gamma$ & 6 & 6 & 0\tabularnewline
$\Gamma$ & 13 & 107 & 96\tabularnewline
\end{tabular}
\par\end{centering}
\caption{Storage costs for the $t$ and $\lambda$ amplitudes, one-electron
integrals $f$, two-electron integrals $v$, and one- and two-particle
density matrices $\gamma$ and $\Gamma$, respectively, that appear
in unrestricted PP, PQ, and PH. PP includes all rank-1 subtensors,
PQ includes all rank-1 and rank-2 subtensors and PH includes all rank-1
through rank-3 subtensors.\label{tab:Storage-costs-for}}
\end{table}

\begin{table}
\begin{centering}
\begin{tabular}{cccc}
 & rank-1 & rank-2 & rank-3\tabularnewline
\hline 
\hline 
$T_{1}/\Lambda_{1}$ & 2 & 2 & \tabularnewline
$T_{2}/\Lambda_{2}$ & 1 & 9 & 8\tabularnewline
$T_{3}/\Lambda_{3}$ &  & 5 & 29\tabularnewline
$T_{4}/\Lambda_{4}$ &  & 1 & 21\tabularnewline
$T_{5}/\Lambda_{5}$ &  &  & 5\tabularnewline
$T_{6}/\Lambda_{6}$ &  &  & 1\tabularnewline
\end{tabular}
\par\end{centering}
\caption{Number of amplitude subtensors at different levels of truncation,
including all spin blocks. PP includes all rank-1 subtensors, PQ includes
all rank-1 and rank-2 subtensors and PH includes all rank-1 through
rank-3 subtensors.\label{tab:noamps}}
\end{table}

\section{Implementation}

We have written a C++ program that generates C++ source code for the
PP, PQ, and PH models, starting from CC equations in spin-orbital
form. In the generated code, the subtensors are stored in memory separately
as arrays. Subtensor symmetry, like $t_{PQ}^{PQ}(P,Q)=t_{PQ}^{PQ}(Q,P)$
for the same-spin doubles excitation amplitudes, is not used in the
storage. As is usual for CC theory, the implementation bases on the
use of intermediates in evaluating nested tensor products, such as
in the singles contribution to the doubles amplitude 
\begin{align*}
t_{ij}^{ab}\leftarrow & \sum_{klcd}t_{k}^{a}t_{l}^{b}t_{i}^{c}t_{j}^{d}v_{kl}^{cd}.
\end{align*}
Pairings are also generated for the intermediates, and the index antisymmetries
of the intermediates are used to eliminate labelings vanishing by
Pauli exclusion. The truncation of the intermediates and integrals
is controlled separately from that of the excitation amplitudes. For
PQ, the intermediates can be truncated to either 2 or 3 pairs, while
for PH a 3-pair truncation is always used. As a result, the solution
of PQ will scale as $O(N^{3})$ or $O(N^{4})$, depending on which
truncation is used, and PH will also scale as $O(N^{4})$. In the
present work, both PQ and PH use 3-pair intermediates, yielding asymptotic
$O(N^{4})$ scaling for both models.

The program performs the pairings by \emph{brute force} by looping
over all the necessary index permutations of the amplitude diagrams
in spin-orbital form, performing the integration over spin into different
spin blocks, and performing the pairing for each block separately.
The generation of the PQ model was found to take roughly 3 minutes
on a single core of a workstation, while generation of the PH model
took 14.5 days on 32 cores on a cluster node. Almost all the time
for the PH model was spent in pairing the hextuple $t$ excitation
amplitudes.

The equations for the subtensor contributions are saved on disk in
human-readable form. OpenMP parallellization is performed over the
individual contributions. For example, the PQ model has 4529 different
contributions to the $t$ amplitudes, while PH has 213234 different
contributions to the $t$ amplitudes. Altogether, the generated PQ
equations take 27 megabytes of disk space in 10080 files, while the
PH equations take four gigabytes of disk space in almost a quarter
million files.

Besides merging contributions that require identical sets of intermediates,
no attempt has been made to search for an optimal reduced set of contributions.
This procedure reduces the amount of separate contributions by roughly
one half from the numbers given above. While common subexpression
elimination techniques commonly performed in conventional CC theory
could clearly be used here, it appears that because of the introduction
of a large variety of subtensors, more optimal sets of intermediates
could be found with a thorough graph search. 

Only a single copy of the integrals is kept in memory. A disk based
direct inversion in the iterative subspace\citep{Pulay1980} (DIIS)
approach is used to accelerate the convergence of the CC amplitudes\citep{Scuseria1986}.

In contrast to conventional CC theory, the pairing of the indices
introduces elementwise (Hadamard) procucts in the pairwise tensor
products. Now, while the number of permutations in a rank-$n$ tensor
increases as $n!$, the number of possible products of two rank-$n$
tensors would increase as $(n!)^{2}$, which becomes unmanageable
even for relatively small $n$. For this reason, our implementation
is based on hardcoded C++ computation kernels for each permutationally
invariant type of contribution, where indices are either fixed (output
index appears only on one of the two tensors), elementwise multiplied
over (output index appears on both tensors), or contraction indices
(index appears on both tensors but is not an output index). The kernels
are machine generated in an external library, both as a \emph{for}-loop-only
implementation, and as code calling basic linear algebra subprogram
(BLAS) routines wherever applicable. When the library is generated,
the two kinds of implementations are benchmarked to determine which
one to use for a given kernel. 

The two-electron integrals are evaluated in the program from B matrices
that can be generated by the resolution of the identity\citep{Vahtras1993}
(RI) or Cholesky decomposition\citep{Beebe1977} (CD) techniques,
as their integral transforms scale more favorably than that of the
traditional two-electron integrals, and as the RI/CD representation
allows for cherrypicking of the wanted integral elements. The Fock
and RI/CD B matrices may currently be obtained from either Q-Chem\citep{Epifanovsky2013a,Shao2014}
or ERKALE\citep{Lehtola2012,erkale}. Both programs have been utilized
in the present work: Q-Chem for calculations using PP orbitals, and
ERKALE for calculations with localized Hartree–Fock orbitals. In the
latter case, the active occupied orbitals are localized using the
generalized Pipek–Mezey method\citep{Lehtola2014} using the intrinsic
atomic orbital partial charge estimate\citep{Knizia2013}. The gradient
descent method used for the localization has been described elsewhere\citep{Lehtola2013a}.
After the occupied space has been localized, corresponding virtual
orbitals are obtained for each active occupied orbital using the Sano
procedure\citep{Sano2000}. All calculations performed with Q-Chem
used exact integrals for the formation of the Fock matrices, and RI
or CD for forming the B matrices for the two-electron integrals for
the CC procedure. The calculations performed with ERKALE used only
Cholesky integrals for both the Fock and B matrices. To ensure near-exactness,
a very small truncation threshold ($10^{-8}$) was used for the Cholesky
procedure both in ERKALE and Q-Chem.

\section{Results}

\subsection{Scaling}

To demonstrate the scaling of the novel implementations of PQ and
PH with respect to the original implementation of the PQ model, we
study full-valence calculations on linear all-trans polyenes \ce{C_nH_{n+2}}
with the geometries described in reference \citenum{Hachmann2006}.
The STO-3G basis set\citep{Hehre1969} and PP optimized orbitals are
used, and the two-electron integrals are formed with RI using the
auxiliary basis set corresponding to the correlation consistent double-$\zeta$
basis set\citep{Weigend2002a}. The timings shown correspond to the
use of a single core on one of the authors' (S.L.) desktop computer
with an Intel Core i7-4770 processor and 32 gigabytes of memory. No
data are shown for sparse PH, because due to inefficiencies in its
implementation getting sensible scaling data would have required hundreds
of gigabytes of memory.

\begin{figure}
\begin{centering}
\includegraphics[width=\columnwidth]{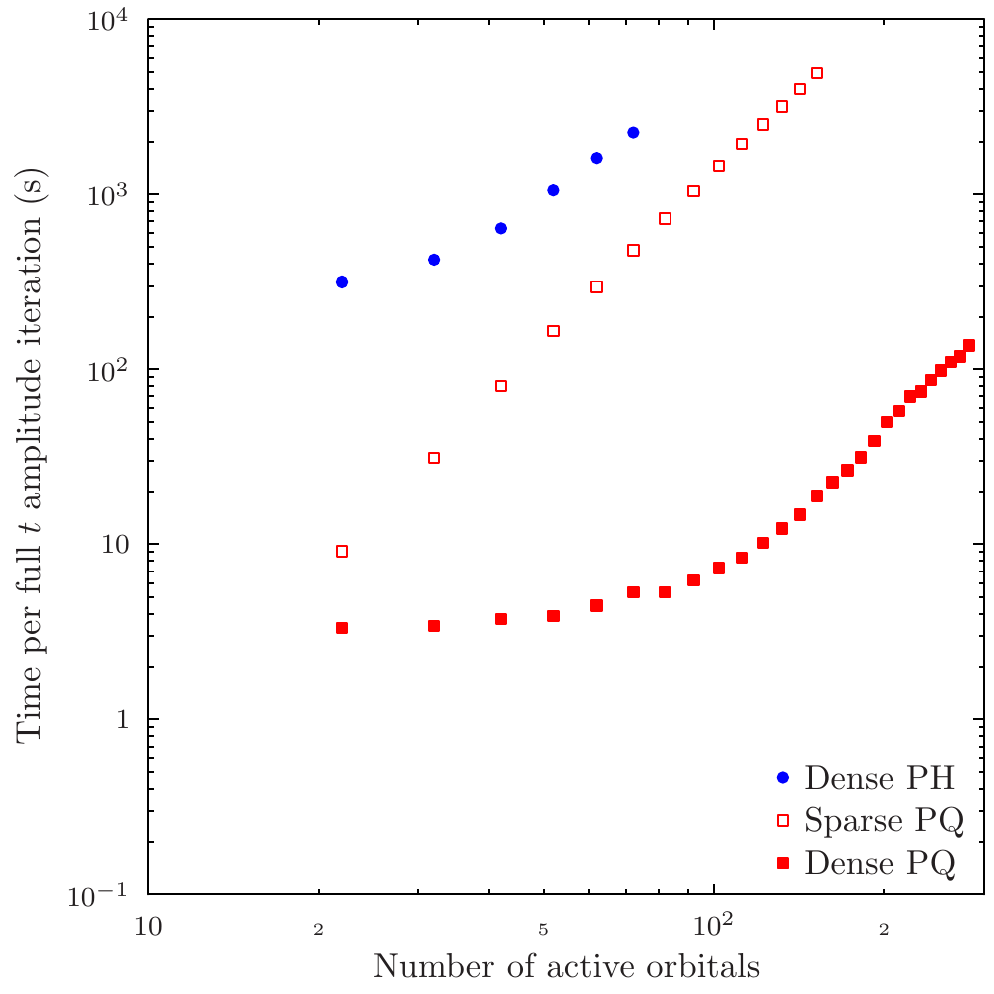}
\par\end{centering}
\caption{Scaling of the new dense tensor implementations of PQ and PH compared
to the earlier implementations\citep{Parkhill2010a} based on sparse
tensors. Note the logarithmic scale.\label{fig:Scaling-of-the}}
\end{figure}

While the use of local PP orbitals in a quasi-1D molecule clearly
would favor algorithms employing sparsity, the results shown in \figref{Scaling-of-the}
emphatically show the tremendous speedup achieved with the dense tensor
formalism that does not use screening of small elements at all. The
distribution of the computational work with the PQ model is shown
in \tabref{Distribution-of-computational}. As is seen, most of the
work in the model is performed in the multiplications, with permutations
also contributing a nontrivial fraction of the total runtime. The
compute kernels and permutations account altogether for 85\% of the
runtime of the model, with another 11\% being spent on dynamic memory
management. A small fraction of the total runtime for large systems
is lost due to inefficiencies in the present implementation, which
are also obvious in \figref{Scaling-of-the} for small systems for
which the time for solution is quasi-constant. (Due to the larger
size of the equations, the inefficiencies are even larger for PH.)
The scaling plot also shows some bumps, which are likely artifacts
of background processes running simultaneously on the workstation. 

The mathematical scaling of the models is determined by the individual
multiplication kernels that appear in the model. The individual multiplication
kernels for the three-pair intermediate PQ model are shown in \tabref{Multiplication-kernels-that}.
Here, all the kernels that contain a contraction use BLAS implementations.
For instance, the heaviest kernel in \tabref{Multiplication-kernels-that}
is implemented with a \emph{dgemv} call within a \emph{for}-loop over
$x$, over which there is an elementwise product. Similarly, the second-heaviest
kernel is also a \emph{dgemv} call inside a \emph{for}-loop over the
$y$ index. The $7^{\text{th}}$ heaviest kernel is just a single
\emph{dgemm }call. As can be seen from the kernels, PQ on \ce{C52H54}
with an active space of 202 electrons in 202 orbitals is still in
the cubic scaling regime, as the heaviest kernels on the list are
cubic scaling. In contrast, PH on \ce{C14H16} with an active space
of 72 electrons in 72 orbitals is on the verge of turning from cubic
to quartic scaling, as while the heaviest kernel (3.3 processor hours)
is cubic scaling, the second-heaviest kernel (2.7 processor hours)
is quartic scaling.

The novel implementation of PQ is fast even for large orbital spaces,
and typically much greater effort is spent in the orbital optimization
with either the self-consistent field method or PP, as well as the
integral transforms: detailed timings for the different steps in the
calculation are shown in \tabref{Comparison-of-timings}. As can be
seen from \figref{Scaling-of-the}, the PQ calculation for an active
space of 140 electrons in 140 orbitals can be performed in a matter
of minutes on a single core with the current version of the code that
has not yet been heavily optimized. Even lower scaling could also
be obtained by truncating the intermediates in PQ to two pairs, but
this would also limit the accuracy of the model\citep{Parkhill2009}.

\begin{table*}
\begin{centering}
\begin{tabular}{lr@{\extracolsep{0pt}.}lr@{\extracolsep{0pt}.}lr@{\extracolsep{0pt}.}lr@{\extracolsep{0pt}.}lr@{\extracolsep{0pt}.}lr@{\extracolsep{0pt}.}lr@{\extracolsep{0pt}.}lr@{\extracolsep{0pt}.}lr@{\extracolsep{0pt}.}lr@{\extracolsep{0pt}.}lr@{\extracolsep{0pt}.}lr@{\extracolsep{0pt}.}l}
Molecule & \multicolumn{2}{c}{$n_{\text{orb}}$} & \multicolumn{2}{c}{SCF} & \multicolumn{2}{c}{PP} & \multicolumn{2}{c}{B matrix} & \multicolumn{2}{c}{PQ ints} & \multicolumn{2}{c}{PQ t} & \multicolumn{2}{c}{PQ perm} & \multicolumn{2}{c}{PQ flop} & \multicolumn{2}{c}{PH ints} & \multicolumn{2}{c}{PH t} & \multicolumn{2}{c}{PH perm} & \multicolumn{2}{c}{PH flop}\tabularnewline
\hline 
\hline 
\ce{C4H6} & \multicolumn{2}{c}{ 22} & 0&23 & 0&35 & 0&68 & 0&00 & 66&63 & 1&3 & 1&8 & 0&05 & 6942&94 & 4&8 & 14&2\tabularnewline
\ce{C6H8} & \multicolumn{2}{c}{ 32} & 0&27 & 0&73 & 0&86 & 0&01 & 75&42 & 1&8 & 3&9 & 0&15 & 12228&76 & 8&8 & 28&2\tabularnewline
\ce{C8H10} & \multicolumn{2}{c}{ 42} & 0&35 & 1&45 & 1&13 & 0&03 & 90&27 & 2&7 & 7&5 & 0&39 & 20436&82 & 12&8 & 41&9\tabularnewline
\ce{C10H12} & \multicolumn{2}{c}{ 52} & 0&44 & 2&37 & 1&65 & 0&06 & 105&08 & 4&3 & 12&7 & 0&79 & 39056&60 & 16&8 & 48&9\tabularnewline
\ce{C12H14} & \multicolumn{2}{c}{ 62} & 0&64 & 3&65 & 2&47 & 0&11 & 120&68 & 5&6 & 17&8 & 1&49 & 59499&11 & 19&2 & 54&5\tabularnewline
\ce{C14H16} & \multicolumn{2}{c}{ 72} & 1&10 & 5&05 & 3&89 & 0&18 & 144&63 & 7&1 & 21&7 & 2&72 & 83323&67 & 20&4 & 58&4\tabularnewline
\ce{C16H18} & \multicolumn{2}{c}{ 82} & 1&71 & 7&10 & 5&89 & 0&26 & 144&47 & 9&2 & 29&8 & \multicolumn{2}{c}{} & \multicolumn{2}{c}{} & \multicolumn{2}{c}{} & \multicolumn{2}{c}{}\tabularnewline
\ce{C18H20} & \multicolumn{2}{c}{ 92} & 2&48 & 9&92 & 8&83 & 0&37 & 169&61 & 11&9 & 34&9 & \multicolumn{2}{c}{} & \multicolumn{2}{c}{} & \multicolumn{2}{c}{} & \multicolumn{2}{c}{}\tabularnewline
\ce{C20H22} & \multicolumn{2}{c}{102} & 3&54 & 13&07 & 13&56 & 0&48 & 197&28 & 13&7 & 39&6 & \multicolumn{2}{c}{} & \multicolumn{2}{c}{} & \multicolumn{2}{c}{} & \multicolumn{2}{c}{}\tabularnewline
\ce{C22H24} & \multicolumn{2}{c}{112} & 6&98 & 19&06 & 18&81 & 0&68 & 225&59 & 15&5 & 44&0 & \multicolumn{2}{c}{} & \multicolumn{2}{c}{} & \multicolumn{2}{c}{} & \multicolumn{2}{c}{}\tabularnewline
\ce{C24H26} & \multicolumn{2}{c}{122} & 7&83 & 24&13 & 26&78 & 0&89 & 274&37 & 18&5 & 47&5 & \multicolumn{2}{c}{} & \multicolumn{2}{c}{} & \multicolumn{2}{c}{} & \multicolumn{2}{c}{}\tabularnewline
\ce{C26H28} & \multicolumn{2}{c}{132} & 9&11 & 30&51 & 37&99 & 1&11 & 332&18 & 20&6 & 50&0 & \multicolumn{2}{c}{} & \multicolumn{2}{c}{} & \multicolumn{2}{c}{} & \multicolumn{2}{c}{}\tabularnewline
\ce{C28H30} & \multicolumn{2}{c}{142} & 10&53 & 37&52 & 55&58 & 1&40 & 400&48 & 22&5 & 52&8 & \multicolumn{2}{c}{} & \multicolumn{2}{c}{} & \multicolumn{2}{c}{} & \multicolumn{2}{c}{}\tabularnewline
\ce{C30H32} & \multicolumn{2}{c}{152} & 12&07 & 46&01 & 70&23 & 1&73 & 510&74 & 26&5 & 51&5 & \multicolumn{2}{c}{} & \multicolumn{2}{c}{} & \multicolumn{2}{c}{} & \multicolumn{2}{c}{}\tabularnewline
\ce{C32H34} & \multicolumn{2}{c}{162} & 13&64 & 55&16 & 104&76 & 2&11 & 610&46 & 27&1 & 53&4 & \multicolumn{2}{c}{} & \multicolumn{2}{c}{} & \multicolumn{2}{c}{} & \multicolumn{2}{c}{}\tabularnewline
\ce{C34H36} & \multicolumn{2}{c}{172} & 15&62 & 66&35 & 130&54 & 2&51 & 713&33 & 27&6 & 54&9 & \multicolumn{2}{c}{} & \multicolumn{2}{c}{} & \multicolumn{2}{c}{} & \multicolumn{2}{c}{}\tabularnewline
\ce{C36H38} & \multicolumn{2}{c}{182} & 17&26 & 77&59 & 161&94 & 2&99 & 847&72 & 28&5 & 56&1 & \multicolumn{2}{c}{} & \multicolumn{2}{c}{} & \multicolumn{2}{c}{} & \multicolumn{2}{c}{}\tabularnewline
\ce{C38H40} & \multicolumn{2}{c}{192} & 19&32 & 94&63 & 242&02 & 3&50 & 1055&37 & 31&7 & 55&3 & \multicolumn{2}{c}{} & \multicolumn{2}{c}{} & \multicolumn{2}{c}{} & \multicolumn{2}{c}{}\tabularnewline
\ce{C40H42} & \multicolumn{2}{c}{202} & 21&33 & 107&16 & 339&07 & 4&14 & 1351&42 & 30&8 & 51&7 & \multicolumn{2}{c}{} & \multicolumn{2}{c}{} & \multicolumn{2}{c}{} & \multicolumn{2}{c}{}\tabularnewline
\ce{C42H44} & \multicolumn{2}{c}{212} & 25&34 & 125&60 & 436&98 & 4&77 & 1566&79 & 32&0 & 51&7 & \multicolumn{2}{c}{} & \multicolumn{2}{c}{} & \multicolumn{2}{c}{} & \multicolumn{2}{c}{}\tabularnewline
\ce{C44H46} & \multicolumn{2}{c}{222} & 27&52 & 143&47 & 586&43 & 5&48 & 1888&20 & 32&5 & 51&4 & \multicolumn{2}{c}{} & \multicolumn{2}{c}{} & \multicolumn{2}{c}{} & \multicolumn{2}{c}{}\tabularnewline
\ce{C46H48} & \multicolumn{2}{c}{232} & 30&12 & 166&73 & 1108&01 & 6&28 & 2020&66 & 32&0 & 51&7 & \multicolumn{2}{c}{} & \multicolumn{2}{c}{} & \multicolumn{2}{c}{} & \multicolumn{2}{c}{}\tabularnewline
\ce{C48H50} & \multicolumn{2}{c}{242} & 33&16 & 188&38 & 1366&80 & 7&31 & 2343&37 & 32&4 & 51&6 & \multicolumn{2}{c}{} & \multicolumn{2}{c}{} & \multicolumn{2}{c}{} & \multicolumn{2}{c}{}\tabularnewline
\end{tabular}
\par\end{centering}
\caption{Comparison of wall timings in seconds for the computational procedure
of PQ and PH as a function of active orbitals $n_{\text{orb}}$. The
third through sixth columm detail the base procedure: SCF solution
followed by the PP orbital optimization and the formation of the RI
B matrix, while columns 6, 7, 8 and 9 (10, 11, 12 and 13) give the
time needed to compute the integrals for PQ (PH), the total time for
the $t$ amplitude iterations needed to solve PQ (PH), as well as
the fraction of total runtime in percent spent in permutation operations
and floating point compute kernels, respectively. \label{tab:Comparison-of-timings}}
\end{table*}

\begin{table}
\begin{centering}
\begin{tabular}{lr@{\extracolsep{0pt}.}l}
Time spent in multiplication kernels (\tabref{Multiplication-kernels-that}) & 3617&84\tabularnewline
Time spent in addition kernels & 42&34\tabularnewline
Time spent in permutations & 2331&36\tabularnewline
Memory allocation & 761&73\tabularnewline
Integrals & 9&27\tabularnewline
Diagonal zeroing & 39&72\tabularnewline
Denominator application & 0&17\tabularnewline
\hline 
\hline 
Sum of above & 6802&40\tabularnewline
\hline 
Total elapsed wall time & 7047&52\tabularnewline
\end{tabular}
\par\end{centering}
\caption{Distribution of computational effort in the solution of the PQ model
for \ce{C52H54}. The timings include all steps in solving the PQ
model: the solution of the $t$ and $\lambda$ amplitudes as well
as the formation of the one- and two-particle density matrices $\gamma$
and $\Gamma$, respectively.\label{tab:Distribution-of-computational}}
\end{table}

\begin{table}
\begin{centering}
\begin{tabular}{lr@{\extracolsep{0pt}.}lr@{\extracolsep{0pt}.}lr@{\extracolsep{0pt}.}l}
Kernel & \multicolumn{2}{c}{Time (s)} & \multicolumn{2}{c}{Scaling} & \multicolumn{2}{c}{BLAS}\tabularnewline
\hline 
\hline 
$o(x)\leftarrow l(x)r(x)$ & 0&00 & \multicolumn{2}{c}{$N$} & \multicolumn{2}{c}{no}\tabularnewline
$o(x,y)\leftarrow l(x,y)r(x,y)$ & 0&01 & \multicolumn{2}{c}{$N^{2}$} & \multicolumn{2}{c}{no}\tabularnewline
$o(x)\leftarrow l(x)r(x)$ & 0&05 & \multicolumn{2}{c}{$N$} & \multicolumn{2}{c}{no}\tabularnewline
$o(x,y)\leftarrow l(x)r(y)$ & 0&07 & \multicolumn{2}{c}{$N^{2}$} & \multicolumn{2}{c}{no}\tabularnewline
$o(x)\leftarrow\sum_{y}l(x,y)r(y)$ & 0&56 & \multicolumn{2}{c}{$N^{2}$} & \multicolumn{2}{c}{yes}\tabularnewline
$o(x,y,z)\leftarrow l(y,z)r(x)$ & 1&27 & \multicolumn{2}{c}{$N^{3}$} & \multicolumn{2}{c}{no}\tabularnewline
$o(x,y)\leftarrow l(x,y)r(y)$ & 13&51 & \multicolumn{2}{c}{$N^{2}$} & \multicolumn{2}{c}{no}\tabularnewline
$o(x)\leftarrow\sum_{y}l(x,y)r(x,y)$ & 21&25 & \multicolumn{2}{c}{$N^{2}$} & \multicolumn{2}{c}{yes}\tabularnewline
$o(x,y)\leftarrow l(x,y)r(x)$ & 23&31 & \multicolumn{2}{c}{$N^{2}$} & \multicolumn{2}{c}{no}\tabularnewline
$o(x,y)\leftarrow\sum_{z}l(x,z)r(y,z)$ & 24&90 & \multicolumn{2}{c}{$N^{3}$} & \multicolumn{2}{c}{yes}\tabularnewline
$o(x,y,z)\leftarrow l(x,y,z)r(z)$ & 28&69 & \multicolumn{2}{c}{$N^{3}$} & \multicolumn{2}{c}{no}\tabularnewline
$o(x,y)\leftarrow l(x,y)r(x,y)$ & 47&89 & \multicolumn{2}{c}{$N^{2}$} & \multicolumn{2}{c}{no}\tabularnewline
$o(x,y,z)\leftarrow\sum_{w}l(y,z,w)r(x,w)$ & 83&04 & \multicolumn{2}{c}{$N^{4}$} & \multicolumn{2}{c}{yes}\tabularnewline
$o(x,y,z)\leftarrow l(x,z)r(y,z)$ & 110&31 & \multicolumn{2}{c}{$N^{3}$} & \multicolumn{2}{c}{no}\tabularnewline
$o(x,y,z)\leftarrow l(x,y,z)r(x)$ & 125&27 & \multicolumn{2}{c}{$N^{3}$} & \multicolumn{2}{c}{no}\tabularnewline
$o(x,y,z)\leftarrow\sum_{w}l(x,y,w)r(z,w)$ & 127&99 & \multicolumn{2}{c}{$N^{4}$} & \multicolumn{2}{c}{yes}\tabularnewline
$o(x,y,z)\leftarrow l(x,y,z)r(x,z)$ & 145&91 & \multicolumn{2}{c}{$N^{3}$} & \multicolumn{2}{c}{no}\tabularnewline
$o(x,y,z)\leftarrow l(x,y,z)r(x,y)$ & 151&28 & \multicolumn{2}{c}{$N^{3}$} & \multicolumn{2}{c}{no}\tabularnewline
$o(x,y,z)\leftarrow l(x,y)r(y,z)$ & 501&45 & \multicolumn{2}{c}{$N^{3}$} & \multicolumn{2}{c}{no}\tabularnewline
$o(x,y,z)\leftarrow l(x,y)r(x,z)$ & 502&06 & \multicolumn{2}{c}{$N^{3}$} & \multicolumn{2}{c}{no}\tabularnewline
$o(x,y)\leftarrow\sum_{z}l(x,y,z)r(y,z)$ & 521&73 & \multicolumn{2}{c}{$N^{3}$} & \multicolumn{2}{c}{yes}\tabularnewline
$o(x,y)\leftarrow\sum_{z}l(x,y,z)r(x,z)$ & 1187&29 & \multicolumn{2}{c}{$N^{3}$} & \multicolumn{2}{c}{yes}\tabularnewline
\end{tabular}
\par\end{centering}
\caption{Multiplication kernels that appear in the three-pair intermediates
PQ model with timings for the \ce{C52H54} molecule.\label{tab:Multiplication-kernels-that}}
\end{table}

\subsection{Accuracy}

To study the accuracy of the generated models in the perfect pairing
hierarchy, in the following we apply the PP, PQ, and PH models on
the $\pi$ space of the polyenes used above for the scaling study,
the symmetric and asymmetric dissociation of the \ce{H50} chain,
as well as the strongly correlated $\pi$ space of polyacene molecules.
Hydrogen chains and polyacenes are standard chemical models for testing
models of strong correlation\citep{Tsuchimochi2009,Sinitskiy2010,Phillips2014}.
The $\pi$ space correlation energies for the polyenes are shown in
\tabref{Polyene--space}, from which the accuracy of the PP hierarchy
already becomes apparent: even for the largest system PH captures
99.3\% of the full correlation energy.

\begin{table}[h]
\begin{centering}
\begin{tabular}{ccccc}
Molecule & PP & PQ & PH & DMRG\tabularnewline
\ce{C4H6} & -0.080460 & -0.091502 & -0.091502 & -0.091502\tabularnewline
\ce{C8H10} & -0.143334 & -0.172876 & -0.176908 & -0.177127\tabularnewline
\ce{C12H14} & -0.205042 & -0.253097 & -0.261466 & -0.262297\tabularnewline
\ce{C16H18} & -0.266492 & -0.333088 & -0.345899 & -0.347403\tabularnewline
\ce{C20H22} & -0.327882 & -0.413030 & -0.430310 & -0.432498\tabularnewline
\ce{C24H26} & -0.389257 & -0.492961 & -0.514717 & -0.517591\tabularnewline
\ce{C28H30} & -0.450629 & -0.572889 & -0.599123 & -0.602684\tabularnewline
\ce{C32H34} & -0.512000 & -0.652816 & -0.683528 & -0.687777\tabularnewline
\ce{C36H38} & -0.573371 & -0.732743 & -0.767934 & -0.772870\tabularnewline
\ce{C40H42} & -0.634742 & -0.812669 & -0.852338 & -0.857963\tabularnewline
\ce{C44H46} & -0.696112 & -0.892595 & -0.936743 & -0.943056\tabularnewline
\ce{C48H50} & -0.757483 & -0.972521 & -1.021147 & -1.028149\tabularnewline
\end{tabular}
\par\end{centering}
\caption{Polyene $\pi$ space correlation energy in the STO-3G basis set with
generalized Pipek–Mezey localized occupied orbitals paired with Sano
virtuals with the PP, PQ, and PH models, compared to DMRG values from
reference \citenum{Hachmann2006}. The size of the active space coincides
with the number of C atoms in the molecule. Note that PQ and PH are
exact for the (4e,4o) system in \ce{C4H6}.\label{tab:Polyene--space}}
\end{table}

Next, while \ce{H50} is chemically uninteresting, it is an excellent
system for studying models for strong correlation. When the interatomic
separation is increased, the system switches from a metallic state
to an insulating state with strong multireference character in the
intermediate region that isn't captured by CC with full single and
double (CCSD) or CC with full single and double and perturbative triple
excitations (CCSD(T))\citep{Hachmann2006}. In the symmetric dissociation,
the equispaced chain is stretched apart into noninteracting \ce{H}
atoms. In the asymmetric dissociation, the system is composed of 25
\ce{H2} molecules with bond length $R=1.4a_{0}$, where $a_{0}$
is the Bohr radius, that are placed on the $z$ axis, and the energy
is studied as a function of the intermolecular distance. Here, the
STO-6G basis set\citep{Hehre1969} and PP optimized orbitals are used.
The two-electron integrals for PQ and PH are generated with CD using
the threshold $\Delta_{\text{Cholesky}}=10^{-8}$. The DMRG data has
been taken from reference \citenum{Hachmann2006}.

\begin{figure}
\begin{centering}
\includegraphics[width=\columnwidth]{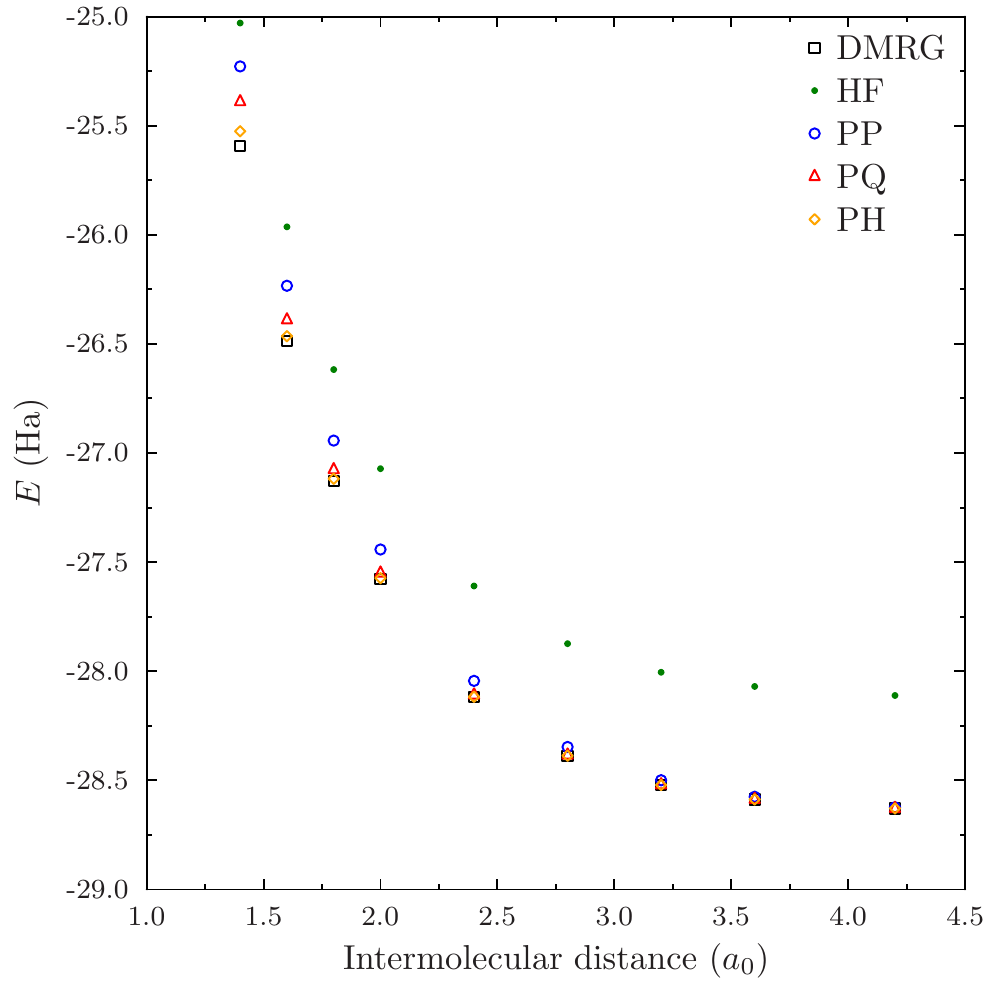}
\par\end{centering}
\caption{Asymmetric dissociation of the \ce{H50} chain into equidistant hydrogen
molecules with fixed bond length $R=1.4a_{0}$.\label{fig:asym}}
\end{figure}

\begin{figure}
\begin{centering}
\includegraphics[width=\columnwidth]{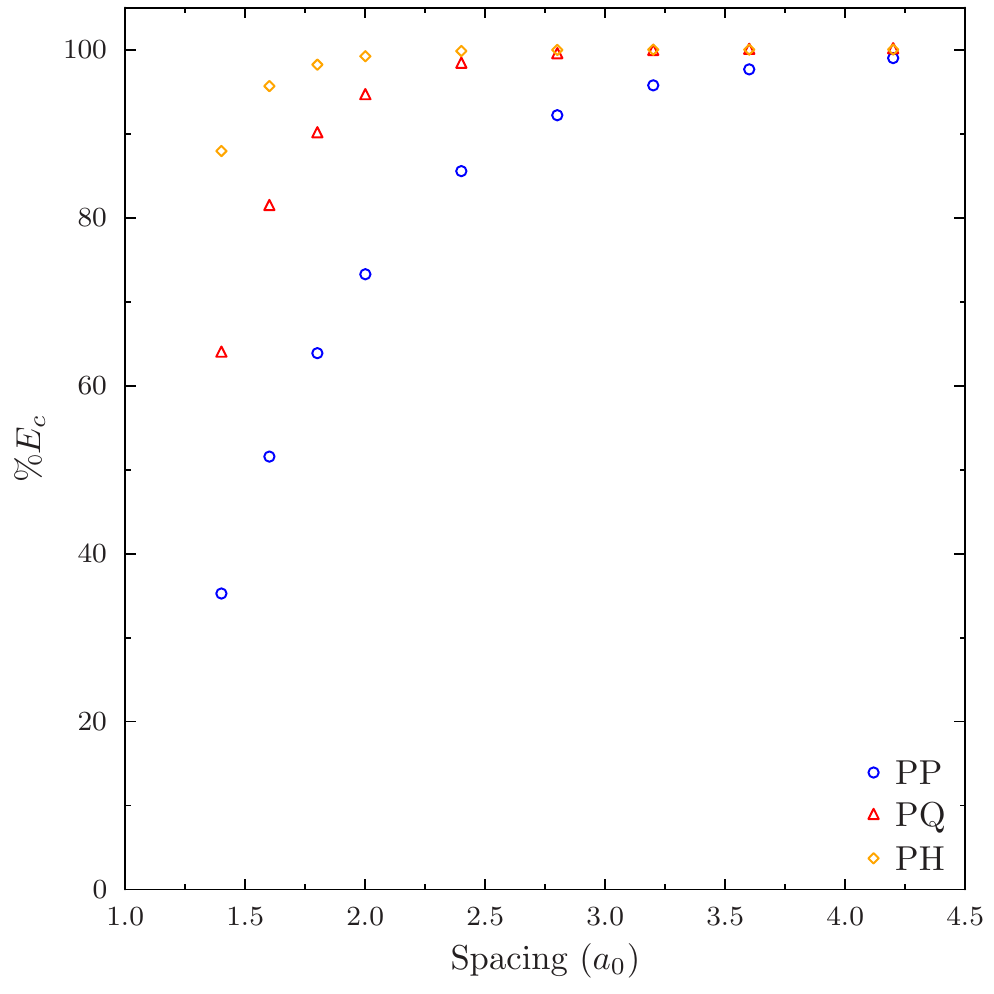}
\par\end{centering}
\caption{Fraction of correlation energy captured by the PP, PQ, and PH models
for the asymmetric dissociation of the \ce{H50} chain.\label{fig:fracasym}}
\end{figure}

The obtained total energies for asymmetric dissociation are shown
in \figref{asym}. For large intermolecular separation, the system
is essentially composed of noninteracting \ce{H2} molecules, for
which already PP is exact. When the molecules are pushed closer, the
differences in the performance of the models become apparent: while
PP quickly starts deviating from the exact solution, PQ and PH stay
more accurate on a wider range of correlation strength. This is also
clearly visible in the fraction of correlation energy captured by
the models shown in \figref{fracasym}. For the most strongly correlated
geometry, in which the atoms are equidistant, PP, PQ, and PH capture
35.2\%, 63.9\% and 87.9\% of the correlation energy, respectively.
PH becomes practically exact (disagreeing from the DMRG correlation
energy by at most 1\%) at the distance $d=2.0a_{0}$, where it captures
99.2\% of the correlation energy. For PQ, practical exactness is reached
at $d=2.8a_{0}$ with 99.4\% of the correlation energy retained. PP
reaches exactness at $d=4.2a_{0}$, where it describes 99.0\% of the
correlation.

\begin{figure}
\begin{centering}
\includegraphics[width=\columnwidth]{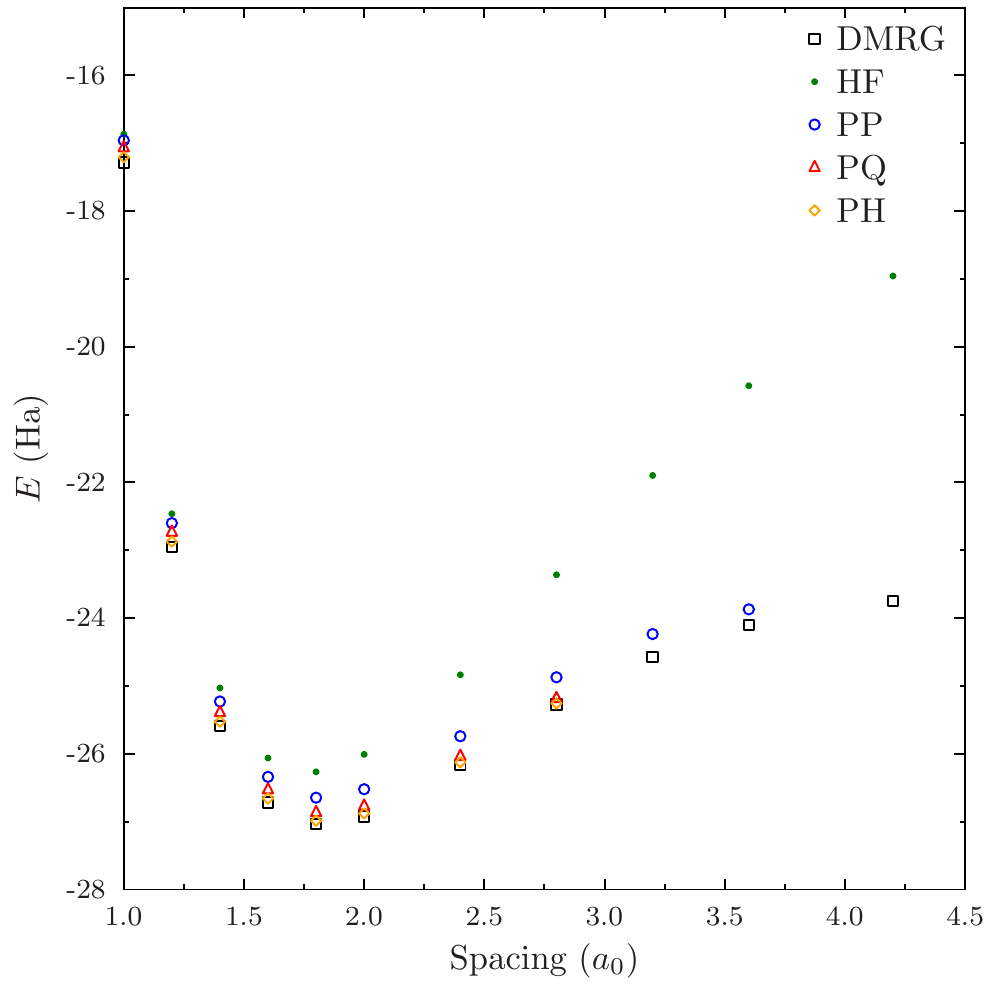}
\par\end{centering}
\caption{Symmetric dissociation of the \ce{H50} chain into equidistant hydrogen
atoms.\label{fig:sym}}
\end{figure}

\begin{figure}
\begin{centering}
\includegraphics[width=\columnwidth]{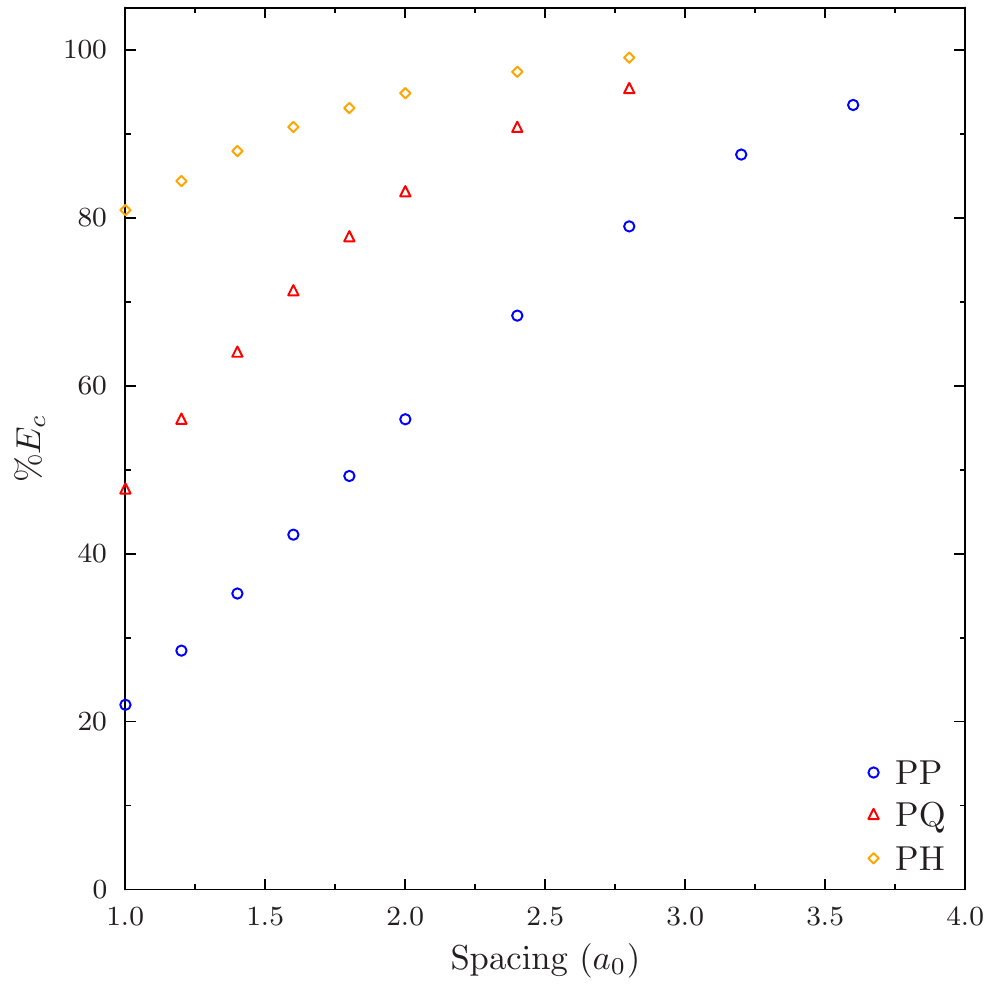}
\par\end{centering}
\caption{Fraction of correlation energy captured by the PP, PQ, and PH models
for the symmetric dissociation of the \ce{H50} chain.\label{fig:fracsym}}
\end{figure}

The symmetric dissociation is a much harder problem, but even here
the models fare seemingly well, as shown by the total and correlation
energy plots in \figref{sym,fracsym}, respectively. As before with
the asymmetric stretch, a clear hierarchy is again seen between the
models, with the most strongly correlated geometry showing 22.0\%,
47.6\%, and 80.9\% of the correlation energy being captured by PP,
PQ, and PH, respectively. As the chain is pulled apart, the models
quickly become more accurate, PH reaching practical exactness at separation
$d=2.8a_{0}$ where it captures 99.0\% of the correlation energy.
Unfortunately, for larger interatomic spacings the CC iterations diverged
for PQ and PH (both with the zero and PP guess for the amplitudes),
thus some data points are missing. However, because PP appears to
become better and better for large separations, the divergence issue
might be solved by the use of a better preconditioner.

\clearpage{}

\begin{figure*}
\subfloat[PP]{\includegraphics[width=0.5\textwidth]{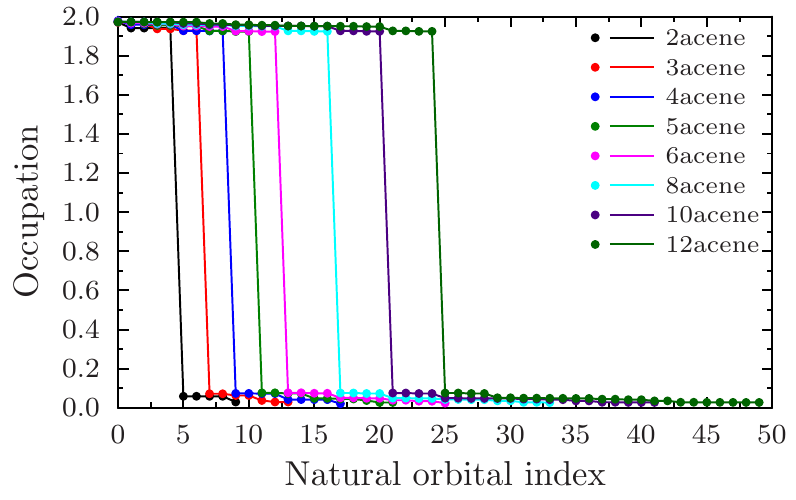}}\subfloat[PQ]{\includegraphics[width=0.5\textwidth]{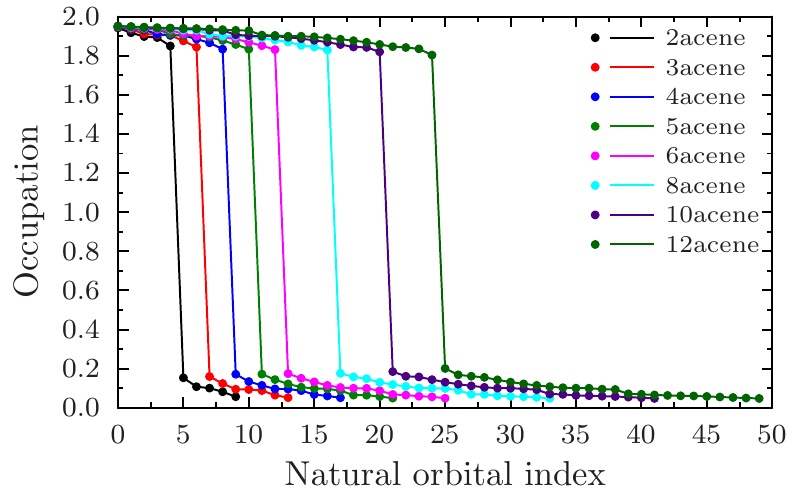}

}

\subfloat[PH]{\includegraphics[width=0.5\textwidth]{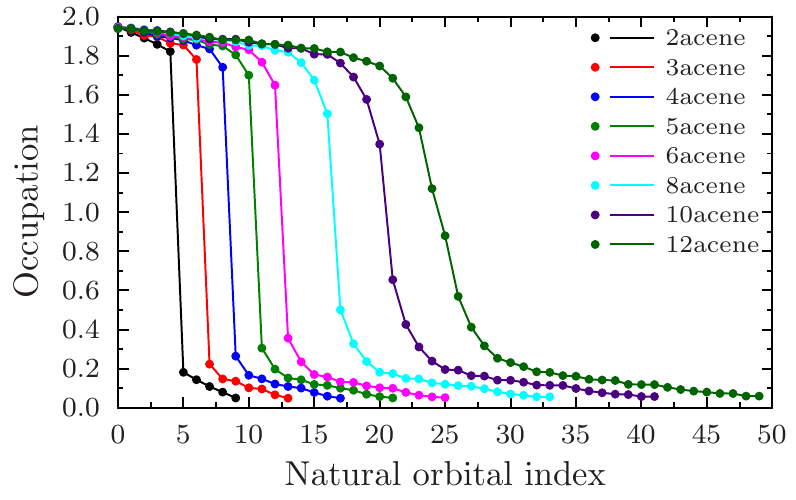}

}\subfloat[DMRG data from reference \citenum{Hachmann2007}]{\includegraphics[width=0.5\textwidth]{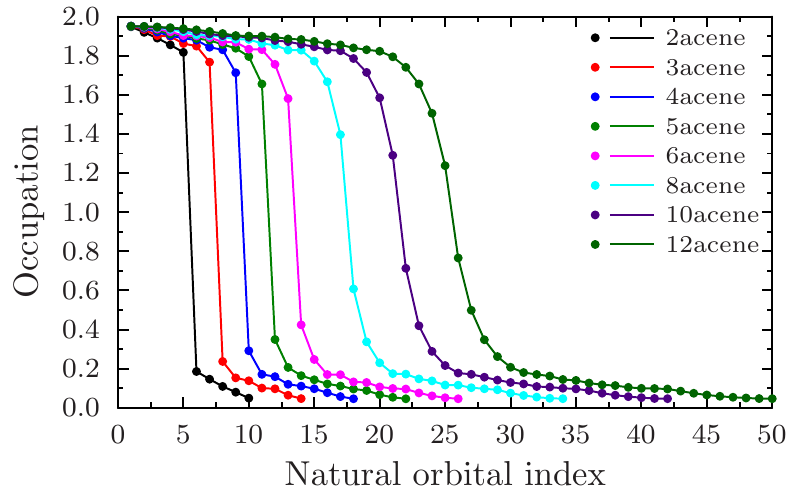}

}

\caption{Polyacene $\pi$ space natural orbital occupation numbers.\label{fig:Polyacene--space}}
\end{figure*}

The linear polyacenes are known to exhibit multiradical behavior in
the $\pi$ space\citep{Hachmann2007}. Next, we will try to match
the DMRG energies in the STO-3G basis from reference \citenum{Hachmann2007}
using their geometries. Generalized Pipek–Mezey localized active occupied
orbitals paired with Sano virtuals were used. The energies obtained
with the PP, PQ, and PH models compared against the DMRG reference
are shown in \tabref{acenes}. Here, the models capture 44\%–59\%,
69\%–88\%, and 96\%–99\% of the DMRG correlation energy for PP, PQ,
and PH, respectively. We can also compare the natural occupation numbers
produced by PP, PQ, and PH against the ones from DMRG; this is done
in \figref{Polyacene--space}. While PP and PQ succesfully capture
a significant amount of the correlation energy, they fail to capture
the strong correlation effects in the $\pi$ space of the polyacenes.
For PP, the occupation number plot has an almost step function like
character, while PQ reproduces some curvature but is still far from
the DMRG reference. In contrast, PH reproduces a clear signal of strong
correlation that is strikingly similar to the DMRG reference values.
Quantitative agreement with DMRG is not, however, reached, which we
primarily attribute to the use of non-optimal orbitals. Indeed, other
choices for the orbital localization method (see the Supplementary
Material) demonstrate that a different choice for the active space
orbitals affects the natural orbital occupation numbers, as well as
the fraction of energy captured. Furthermore, as Pipek–Mezey localized
Hartree–Fock orbitals were found to yield better energies than PP
optimized orbitals at the UPH level of theory, we conclude that as
has been seen for CCVB\citep{Small2014}, PQ and PH presumably need
orbitals that are less localized than those reproduced by PP, and
the best choice for the (non-optimized) orbitals is still an open
question.

\begin{table*}[h]
\begin{centering}
\begin{tabular}{cccccc}
Molecule & Active space & PP & PQ & PH & DMRG\tabularnewline
2acene & (10e,10o) & -0.104902 & -0.157684 & -0.177098 & -0.178294\tabularnewline
3acene & (14e,14o) & -0.144544 & -0.213115 & -0.249895 & -0.254307\tabularnewline
4acene & (18e,18o) & -0.190182 & -0.279345 & -0.325491 & -0.332661\tabularnewline
5acene & (22e,22o) & -0.228839 & -0.335088 & -0.401098 & -0.412627\tabularnewline
6acene & (26e,26o) & -0.265084 & -0.393537 & -0.479533 & -0.495683\tabularnewline
8acene & (34e,34o) & -0.322287 & -0.495219 & -0.641147 & -0.668526\tabularnewline
10acene & (42e,42o) & -0.382622 & -0.595840 & -0.803505 & -0.838580\tabularnewline
12acene & (50e,50o) & -0.444834 & -0.698165 & -0.968521 & -1.007378$^{a}$\tabularnewline
\end{tabular}
\par\end{centering}
\caption{Polyacene $\pi$ space correlation energy in the STO-3G basis set
with generalized Pipek–Mezey localized occupied orbitals paired with
Sano virtuals with the PP, PQ, and PH models, compared to DMRG values
from reference \citenum{Hachmann2007}. Note that PQ and PH are exact
for the (4e,4o) system in \ce{C4H6}. $^{a}$DMRG value of reference
\citenum{Hachmann2007} was not converged.\label{tab:acenes}}
\end{table*}

\section{Summary and Discussion}

We have described the low-scaling dense tensor implementation of a
family of truncated CC methods, and generated efficient implementations
of the PQ and PH models. High-rank tensors are expressed in terms
of a considerable number of subtensors, and summations over Kronecker
delta indices are performed before computer code is generated. As
straightforward truncations of CC theory, the models inherit all the
properties of the parent theory, such as size extensivity, and pre-existing
machinery for the calculation of properties could be used.

We have demonstrated impressive speedups (of roughly 200-500 times)
compared to the earlier implementation that already had the correct
$O(N^{4})$ scaling for PQ, and demonstrated the rapid convergence
of the PP, PQ, and PH models in describing strong correlation with
applications to the dissociation of the \ce{H50} chain. Although
only minimal basis sets have been used in the present work due to
limited availability of high-level reference data, the runtime of
the models after integral transformation is basis set independent
due to the use of an active space (and the lack of orbital optimization
that has not been pursued in the present work). While the models are
already computationally fast enough to allow their use in studies
of chemical problems, we believe they can still be made even faster
by a variety of approaches. Most importantly, the role of the intermediates
in the paired theories are clearly different from the intermediates
in full-rank CC theory. A thorough procedure for identifying common
factors between different contributions, as well as a graph-type approach
for determining the optimal set of intermediates, could plausibly
yield a further order of magnitude speedup to the existing implementation.
In the case of closed-shell molecules, spin summation of the model
would reduce storage costs for the amplitudes by roughly 50\%, and
cut down on computational costs as well. A reorganization of the way
the amplitude updates are performed might yield convergence in a smaller
number of iterations\citep{Matthews2015}. Because of the mathematical
structure of the models, the storage of the tensors is already distributed,
and massively parallel implementations of the models could be pursued.

While we have demonstrated that the truncation hierarchy advocated
in the present work allows for applications to large systems, it is
further important to note that in systems with little static correlation
it is possible to further speed things up. While in PQ and PH the
excitation amplitude tensors have similar storage requirements (within
an order of magnitude) as shown in \tabref{noamps}, the runtime tends
to be dominated by the highest excitations due to the larger amount
of antisymmetrizations necessary in the CC diagrams. Thus, should
the higher excitation operators turn out to be insignificant, disabling
them can yield further speedups. For example, the PH model with only
single and double (and triple and quadruple) excitations enabled is
essentially a local CCSD (CCSDTQ) approach, if the treatment is based
on localized orbitals.

As we have seen in the application to the acene series, the choice
of orbitals may be important for the perfect pairing hierarchy of
models, which we will address in future work. In addition, we wish
to extend the PQ and PH models to open-shell systems, as well as to
include the description of dynamic correlation. The similarities between
the PP truncation hierarchy and local CC methods could also be further
explored.

\section*{Supplementary Material}

See supplementary material for results on the polyacenes with various
choices for the active space orbitals.

\section*{Acknowledgments}

S.L. thanks Evgeny Epifanovsky, David Small, and Joonho Lee for discussions,
and Johannes Hachmann for supplying the DMRG natural occupation numbers
for the polyacenes. This work was supported by the Director, Office
of Basic Energy Sciences, Chemical Sciences, Geosciences, and Biosciences
Division of the U.S. Department of Energy, under Contract No. DE-AC02-05CH11231.
J.P. thanks The University of Notre Dame’s College of Science and
Department of Chemistry and Biochemistry for generous start-up funding.

\end{document}